\input harvmac
\input amssym.tex

\def\newdate{Stony Brook, 2/24/2004}

\def\a{\alpha}
\def\b{\beta}  
\def\g{\gamma}
\def\l{\lambda}

\def\e{\epsilon}
\def\t{\theta}

\def\p{\partial}

\Title{
\vbox{\hbox{YITP-SB-04-07}
}}   
{\vbox{
\centerline{Harmonic Superspaces from Superstrings}
}}  
 
\medskip\centerline
{
P.~A.~Grassi$^{~a,b,}$\foot{pgrassi@insti.physics.sunysb.edu},
and 
P.~van~Nieuwenhuizen$^{~a,}$\foot{vannieu@insti.physics.sunysb.edu}
} 
\medskip   
\centerline{$^{(a)}$ 
{\it C.N. Yang Institute for Theoretical Physics,} }  
\centerline{\it State University of New York at Stony Brook,   
NY 11794-3840, USA}  
\medskip
\centerline{$^{(b)}$ {\it Dipartimento di Scienze,
Universit\`a del Piemonte Orientale,}}
\centerline{\it
C.so Borsalino 54, Alessandria,  15100, ITALY}

\medskip  
\vskip  .5cm  
\noindent

 We derive harmonic superspaces for $N=2,3,4$ SYM theory in 
 four dimensions from superstring theory. The pure spinors in ten dimensions 
 are dimensionally reduced and yield the harmonic coordinates. Two anticommuting BRST charges implement Grassmann analyticity and harmonic analyticity. The string field theory action produces the action and 
 field equations for N=3 SYM theory in harmonic superspace. 

\Date{\newdate}


\lref\MovshevIB{
M.~Movshev and A.~Schwarz,
[hep-th/0311132].
}

\lref\Grassione{  
P.~A.~Grassi, G.~Policastro, M.~Porrati and P.~van Nieuwenhuizen,  
JHEP {\bf 10} (2002) 054, 
[hep-th/0112162]; 
P.~A.~Grassi, G.~Policastro, and P.~van Nieuwenhuizen,  
JHEP {\bf 11} (2002) 004, 
[hep-th/0202123].} 

\lref\GrassiWZW{
P.~A.~Grassi, G.~Policastro and P.~van Nieuwenhuizen,
Nucl.\ Phys.\ B {\bf 676}, 43 (2004)
[hep-th/0307056].
}

\lref\purespinors{
\'E. Cartan, {\it Lecons sur la th\'eorie des spineurs}, 
Hermann, Paris (1937);
C. Chevalley, {\it The algebraic theory of Spinors}, 
Columbia Univ. Press., New York, 1954;
 R. Penrose and W. Rindler, 
{\it Spinors and Space-Time}, Cambridge Univ. Press, Cambridge (1984);
P.~Furlan and R.~Raczka,  
J.\ Math.\ Phys.\  {\bf 26}, 3021 (1985);
P. Budinich and A. Trautman, {\it The spinorial chessboard}, Springer, 
New York (1989);
P.S. Howe, 
Phys. Lett. B258 (1991) 141, Addendum-ibid.B259 (1991) 51; 
P.S. Howe,
Phys. Lett. B273 (1991) 
90.}  



\newsec{Introduction}

Pure spinors \purespinors\ in ten 
dimensions are complex commuting 
chiral spinorial ghosts $\l^{\hat\a}$ with $\hat\a=1,\dots,16$ satisfying the 
ten nonlinear constraints 
\eqn\uno{
\l^{\hat\a} \g^{\hat m}_{\hat\a \hat\b} \l^{\hat \b}=0\,,
}
 (hats denote 10-dimensional  indices). 
 They form the starting point for a new approach to the quantization of the superstring 
 with coordinates $x^{\hat m}, \t^{\hat \a}$ and $\l^{\hat \a}$ 
  \lref\berko{  
N.~Berkovits,  
JHEP { 0004}, 018 (2000); N.~Berkovits,
JHEP {\bf 0109}, 016 (2001)
[hep-th/0105050].
N.~Berkovits,
Int.\ J.\ Mod.\ Phys.\ A {\bf 16}, 801 (2001); 
N.~Berkovits,
{\it ICTP lectures on covariant quantization of the superstring,}
hep-th/0209059.
}
\berko. 
 Due to these constraints on $\l$, the troublesome second class constraints 
 of the superstring become effectively first class. One can relax these constraints and obtain a  covariant formulation by introducing more ghosts as Lagrange multipliers \Grassione. The result is an $N=2$ WZNW model \GrassiWZW. 
 The pure spinors in this covariant approach are 
 real and the BRST charge maps $\t^{\hat \a}$ into $\l^{\hat \a}$. In this letter, though, we 
 use complex constrained $\l^{\hat \a}$. Pure spinors also exist in other dimensions 
 \purespinors. 
 
\lref\GalperinUW{
A.~S.~Galperin, E.~A.~Ivanov, V.~I.~Ogievetsky and E.~S.~Sokatchev,
{\it Harmonic Superspace,} Cambridge Univ. Press, 2001. 
A. Galperin, E. Ivanov, S. Kalitzin, V. Ogievetsky and
E. Sokatchev,  Class. Quant. Grav. {\bf 1}, 469 (1984);
A. Galperin, E. Ivanov, S. Kalitzin, V. Ogievetsky and 
Sokatchev, Class. Quant. Grav. {\bf 2}, 155 (1985). 
}

Harmonic superspace (see \GalperinUW\ for a complete 
review of the subject and references\foot{
Two useful accounts of the subject 
can be found in  
\lref\HoweMD{
P.~S.~Howe and G.~G.~Hartwell,
Class.\ Quant.\ Grav.\  {\bf 12}, 1823 (1995).
} \HoweMD\ and in 
\lref\HartwellRP{
G.~G.~Hartwell and P.~S.~Howe,
Int.\ J.\ Mod.\ Phys.\ A {\bf 10}, 3901 (1995)
[hep-th/9412147].
} \HartwellRP. 
Projective harmonic superspace has been introduced in
\lref\projective{
A.~Karlhede, U.~Lindstrom and M.~Rocek,
Phys.\ Lett.\ B {\bf 147}, 297 (1984);
S.~J.~Gates, C.~M.~Hull and M.~Rocek,
Nucl.\ Phys.\ B {\bf 248}, 157 (1984).
} \projective. 
The application to the AdS/CFT correspondence 
 is studied in  
 \lref\Ferrara{
S.~Ferrara and E.~Sokatchev,
Lett.\ Math.\ Phys.\  {\bf 52}, 247 (2000)
[hep-th/9912168]\semi
L.~Andrianopoli, S.~Ferrara, E.~Sokatchev and B.~Zupnik,
Adv.\ Theor.\ Math.\ Phys.\  {\bf 3}, 1149 (1999)
[hep-th/9912007].
} \Ferrara, and 
some developments of $N=4$ harmonic superspace for SYM 
can be found in 
\lref\ZupnikUS{
B.~M.~Zupnik,
[hep-th/0308204].
} \ZupnikUS\ and in 
\lref\ZupnikHD{
B.~M.~Zupnik,
Nucl.\ Phys.\ Proc.\ Suppl.\  {\bf 102}, 278 (2001)
[hep-th/0104114].
} \ZupnikHD.} 
)  
 was constructed to circumvent the no-go theorems for 
 a full-fledged superspace description of N-extended supersymmetries (susy). The
 main idea is to let the $R$-symmetry group $U(N)$ (or $SU(N)$ for N=4), 
 which acts on the 
 susy generators, become part of a coset approach.  The generators of $U(N)$ 
 are divided into coset generators with coset coordinates $u$ called harmonic variables, 
 and subgroup generators. Superfields depend not only on $x^{m}$ and 
 half of the 
 $\t^{\a}_{I}, \bar\t^{\dot\a I}$ (with $\a,\dot\a =1,2$ and  $I=1,\dots,N$)
 but also on $u$'s. For $N=2,3,4$ the cosets most often used are
 \eqn\unomezzo{ 
 {SU(2)\over U(1)}, ~~~~~
 {SU(3) \over U(1) \times U(1)}, ~~~~{SU(4)\over 
 S[U(2)\times U(2)]}\,,
 } 
 respectively, although other choices are also possible \HoweMD. 
 
 In this letter we present a derivation of four-dimensional harmonic superspaces 
 from ten-dimensional pure spinors by using ordinary dimensional reduction in 
 which we set the extra six coordinates to zero by hand. 
 The spinors $\l^{\hat\a}$ decompose into $\l^{\a}_{I}$ and $\bar\l^{\dot\a I}$ 
 where $I=1,\dots,4$ is an $SU(4) \sim SO(6)$ index. 
 The main idea is to factorize the pure spinors 
 $\l^{\hat \a}$ into auxiliary variables $\l^{\a}_{a}$ and $\bar\l^{\dot \a}_{a}$ with $a=1,2$, 
 and harmonic variables $u^{a}_{I}$ and $\bar v^{a I}$. 
In this way we factorize the Lorenz group and the internal symmetry group 
$SU(4)$. Using this factorization, the pure spinor constraints turn into constraints on 
 $\l^{\a}_{a}$ and $\bar\l^{\dot \a}_{a}$, and on $u^{a}_{I}$ and $\bar v^{a I}$. 
 
 Contracting the operator $d_{z \hat\a}$ in the BRST charge \berko
 \eqn\duemezzo{
 Q = \oint dz \l^{\hat\a} d_{z \hat\a}\,,
 }
 with the harmonic coordinates leads to eight spinorial 
 covariant derivatives 
 \eqn\quaderA{
d_{\a}^{a} = u_{I}^{a} d^{I}_{\a}\,, ~~~~~~~
\bar d_{\dot \a}^{a} = \bar v^{a I} \bar d_{\dot\a I}\,, 
}
which satisfy the constraints
\eqn\quaderB{
\{ d_{\a}^{a}, d_{\b}^{b} \} = \e_{\a\b} \{ \bar d_{\dot \a}^{a}, \bar d^{\dot \a b} \}\,, 
~~~~~
\{ d_{\a}^{a}, \bar d_{\dot \b}^{b}\} =0\,, 
}
as a consequence of the constraints on $u$ and $\bar v$, 
and in terms of which G(Grassman)-analyticity (dependence on half the 
$\t$'s) of superfields is defined. 
 
 If one does not provide the information that $d_{\a}^{a}$ and 
 $\bar d^{a}_{\dot \a}$ are linear in $u^{a}_{I}$ and $\bar v^{a I}$, one 
 looses information. We therefore construct a second BRST charge 
 which only anticommutes with $Q_{H}$ if $d_{\a}^{a}$ and 
 $\bar d^{a}_{\dot \a}$ are factorized as in \quaderA. It is 
 constructed from the generators of $U(N)$ represented by the 
 following differential 
 operators\foot{The R-symmetry group $SU(4)$ corresponds to the 
 Lorentz generators in the extra dimensions. This 
 suggests  that the second BRST charge 
 might be obtained by dimensional reduction of the BRST charge in ten 
 dimensions, extended to include the  ten dimensional Lorentz generators.
 } 
 \eqn\quaA{
d_{~~a'}^{a} = u_{I}^{a} \p_{u_{I}^{a'}} - \bar u^{I}_{a'} \p_{\bar u^{I}_{a}}\,.
}
Requiring that the vertex operators are annihilated by these BRST charges
should yield the 
field equations of N=4 harmonic superspace. In this letter we work 
out the case of N=3 and obtain
by truncation the field equations of N=3 SYM theory in 
harmonic superspace. 
We end by deducing an action for N=3 SYM theory 
in harmonic superspace from the Chern-Simons action 
for string field theory 
\lref\wichen{ 
E.~Witten,
Nucl.\ Phys.\ B {\bf 268}, 253 (1986);
E.~Witten, 
[hep-th/9207094]. 
} \wichen. 

 The present analysis might provide a link between 
 string theory with pure spinors and recent 
 developments in twistor theory 
\lref\wittwi{
E.~Witten,
hep-th/0312171.
} \wittwi. 
Another interesting aspect not covered in the present letter 
is deformed harmonic superspace 
\lref\defor{ 
S.~Ferrara and E.~Sokatchev,
Phys.\ Lett.\ B {\bf 579}, 226 (2004)
[hep-th/0308021]; 
E.~Ivanov, O.~Lechtenfeld and B.~Zupnik,
hep-th/0308012.
} \defor. 
It would be 
interesting to discover which kind of harmonic superspace one obtains 
for suitable Ramond-Ramond backgrond fields \lref\Ramond{ 
J.~de Boer, P.~A.~Grassi and P.~van Nieuwenhuizen,
Phys.\ Lett.\ B {\bf 574}, 98 (2003)
[hep-th/0302078]; H.~Ooguri and C.~Vafa,
Adv.\ Theor.\ Math.\ Phys.\  {\bf 7}, 53 (2003)
[hep-th/0302109].
} \Ramond.

In a future article we intend to extend these results to the N=4 case and 
construct an action for N=4 SYM theory 
\lref\PREP{P.~A.~Grassi, M. Ro\v cek, and P. van Nieuwenhuizen, 
{in preparation}} \PREP. In particular, this should give 
a conceptually simple derivation of the rather complicated 
measure.\foot{A similar analysis is pursued in 
\lref\MovshevIB{
M.~Movshev and A.~Schwarz,
[hep-th/0311132].
}
\MovshevIB.
}   
 

\newsec{The coordinates of N=4, N=3, and N=2 harmonic superspace from pure spinors}

We substitute the decomposition $\l^{\hat \a} = (\l^{\a}_{I}, \bar\l^{\dot\a I})$ 
into the pure spinor constraints, and use the representation of the matrices 
$\g^{\hat m}_{\hat \a \hat \b}$ given in \lref\har{J.~Harnad and S. Shnider, Comm. \ Math. \ Phys. {\bf 106}, 183 (1986)} \har. In this representation 
the Dirac matrices with $m=0,1,2,3$ are labelled by $\g^{\a\dot\b}$ and 
those for $m=4,\dots,9$ are labelled by $\g^{IJ} = - \g^{JI}$, and 
all matrix elements are expressed in terms of Kronecker delta's and 
the epsilon symbols $\e^{\a\b}, \e^{\dot\a\dot\b}$ and $\e^{IJKL}$. The pure 
spinor constraints decompose then into the following six plus four 
constraints 
\eqn\reduA{
\l^{\a}_{I} \e_{\a\b} \l^{\b }_{J} + {1\over 2} \e_{IJKL} \bar\l^{\dot\a K} \e_{\dot\a\dot\b} 
\bar\l^{\dot\b L} =0 \,, ~~~~~~~\l^{\a}_{I} \bar\l^{\dot\a I} = 0 \,.}
The first relation corresponds to $m=4,\dots,9$ while the second one corresponds to $m=0,1,2,3$. 
To solve these constraints we adopt the following ansatz
\eqn\reduB{
\l^{\a}_{I} =  \l^{\a}_{a} u^{a}_{I}\,, ~~~~~~~ \bar\l^{\dot\a J} =  
\bar\l^{\dot\a}_a \bar v^{a J} \,,
}
where $a=1,2$. The new variables $u^{a}_{I}$ and $\bar v^{a J}$ are complex and commuting.They carry $GL(2,{\bf C})$ and 
$SU(4)$ indices. The spinors 
$\l^{\a}_a, \bar\l^{\dot\a}_{a}$ are also complex and commuting, and 
carry a representation of $SL(2, {\bf C})$ and 
$GL(2,{\bf C})$. In this way, we separate 
the Lorentz group from the internal symmetry group $SU(4)$. 

The decomposition in \reduB\ is left invariant by the gauge transformations 
\eqn\reduC{
u^{a}_{I} \rightarrow M^a_{~~b} u^{b }_{I}\,, ~~~~
~~~ \l^{\a}_a \rightarrow \l^{\a}_b (M^{-1})_{~~a}^{b} \,,}
$$
\bar v^{a J} \rightarrow \bar M^a_{~~b} \bar v^{b J}\,, 
~~~~~~~ \bar\l^{\dot\a}_a \rightarrow 
\bar\l^{\dot\a}_b (\bar M^{-1})_{~~a}^{b} \,,
$$
where $M$ and $\bar M$ are independent $GL(2,{\bf C})$ matrices. 
The factorization \reduB\ 
plus the gauge invariance \reduC~yields 16 complex parameters. To reduce to the usual 11 independent  complex parameters of pure spinors, we further impose the following two covariant constraints 
\eqn\reduD{
u^{a}_{~I} \bar v^{b I} = 0\,, ~~~~~~  
\l^{\a}_{a} \e_{\a\b} \e^{a b} \l^{\b}_{b} + 
\bar\l^{\dot\a}_{a} \e_{\dot\a\dot\b} \e^{a b} \bar\l^{\dot\b}_{b} = 0 \,. 
}
The first one imposes four complex conditions, while the second equation is a single invariant complex condition. 

The first constraint in \reduD\ and the gauge transformations in \reduC\ reduce the 
16 complex components of $u^{a}_{I}$ and $\bar v^{a I}$ to 8 real parameters. 
This is the same number as the number of independent parameters of the coset
${U(4) \over U(2) \times U(2)} = {SU(4) \over  S(U(2) \times U(2)) }$ used in \HartwellRP
(see also \ZupnikHD\ and \Ferrara). The restriction of 
$U(2) \times U(2)$ to the subgroup $S(U(2) \times U(2))$ 
is due to second constraint of \reduD. The latter is 
preserved by the transformations $M$ and $\bar M$ only after 
the identification ${\rm det} M = {\rm det} \bar M$. 

To identify the $SU(4)$ of the coset space, we introduce 
new coordinates 
$u^{a, \dot b}_{I} =(u^{a, \dot 1}_{I}, u^{a, \dot 2}_{I})$ 
where 
\eqn\reduEE{
u^{a, \dot 1}_{I} = u^{a}_{I}\,, ~~~~
u^{a, \dot 2}_{I} = \e^{ab} v_{b I}\,, }
and $v_{b I}= (\bar v^{b I})^{*}$ . The matrix $u_{I}^{(a,\dot b)}$
is a $U(4)$ matrix because the harmonic variables 
$u^{a}_{I}$ and $\bar v^{a I}$ satisfy the constraints \reduD\ and they can 
be normalized as follows, using the gauge transformations \reduC, 
\eqn\norma{ 
u^{a}_{I} \bar u^{I}_{b} = \delta^{a}_{~b}\,, ~~~~~
\bar v^{a I} v_{b I} = \delta^{a}_{~b}\,, 
}
where $\bar u^{I}_{b} = (u^{b}_{I})^{*}$. 

To restrict $U(4)$ to $SU(4)$ we choose the gauge 
\foot{Denoting this relation by 
$N_{IJ} =0$, it is clear that $N_{IJ} \bar v^{a J} =0$ and 
$\e^{IJKL} N_{KL} u_{J}^{a} =0$ due to (2.4). This leaves the 
phase of ${\rm det} \, u^{a\dot b}_{I}$ undetermined. The gauge in 
(2.5) sets this phase to zero.}  
 \eqn\reduE{
u^{a}_{I} \e_{ab} u^{b}_{J} - {1\over 2} \e_{IJKL} \bar v^{a K} \e_{ab} \bar v^{b L} = 0\,.
}
This gauge choice is preserved by $S(U(2) \times U(2))$. 

The normalizations \norma\ 
fix 4 real parameters for each $GL(2,{\bf C})$ in \reduC. The   
remaining 7 real parameters of $GL(2,{\bf C})$ (remaining after the 
identification ${\rm det} M = {\rm 
det} \bar M$), reproduce the subgroup $S(U(2) \times U(2))$.  All equations are covariant 
under this subgroup. Thus the coordinates $u^{A}_{I} \equiv u^{a, \dot a}_{I}$, 
with $A=1,\dots,4$,  parametrize the coset $SU(4) \over S(U(2) \times U(2))$.

Let us turn to N=3 harmonic superspace. 
If we decompose the $\l^{\a}_{I}$'s and the $\bar\l^{\dot\a I}$'s into 
N=3 vectors and N=3 scalars we have $\l^{\a}_{I} = (\l^{\a}_{i}, \psi^\a)$ and 
$\bar\l^{\dot\a I} = (\bar\l^{\dot\a i}, \bar\psi^{\dot\a})$. In that basis, 
the pure spinor constraints in (2.1) become
$$
\l^{\a}_{i} \e_{\a\b} \l^{\b}_{j} + 
\e_{ijk} \bar\l^{\dot\a k} \e_{\dot\a\dot\b} \bar\psi^{\dot\b} =0\,,
$$
$$
\l^{\a}_{i} \e_{\a\b} \psi^{\b} + \e_{ijk} \bar\l^{\dot\a j} \e_{\dot\a\dot\b} \bar\l^{\dot\b k} =0 \,, $$
\eqn\reduG{
\l^{\a}_{i} \bar\l^{\dot\a i} + \psi^{\a} \bar\psi^{\dot\a} = 0 \,.
}
The reduction to the N=3 case is obtained by setting 
$\psi^\a = \bar\psi^{\dot\a} = 0$.  
Inserting this ansatz into the first two equations of 
\reduG, we obtain 
\eqn\reduGG{
\l^{\a}_{i} \e_{\a\b} \l^{\b}_{j}=0\,, ~~~~~~~
\bar\l^{\dot\a j} \e_{\dot\a\dot\b} \bar\l^{\dot\b k} =0 \,,
}
which is equivalent to requiring that all determinants of order 2 
of the matrices $\l^{\a}_{i}$ and $\bar\l^{\dot \a i}$ vanish.\foot{It is well-known  
(and easy to check) that if two of the $2 \times 2$ submatrices have vanishing determinant, so 
does the third. This implies (2.10).} 
This means that the pure spinors can be factorized into
\eqn\neweq{
\l^{\a}_{i} = \l^\a u_i\,, ~~~~~~\bar\l^{\dot\a i}= \bar\l^{\dot \a} \bar v^i
}
and the equations \reduG\ are solved by
\eqn\harcon{
\psi^\a = \bar\psi^{\dot\a} = 0\,,~~~~~~ u_i \bar v^i = 0\,.
}
So for the N=3 case 
no constraint is needed for $\l^\a$ and $\bar\l^{\dot\a}$. Notice that 
the two complex vectors $u_{i}$ and $\bar v^{i}$ are defined up to a gauge transformation
\eqn\gauuv{
u_{i} \rightarrow \rho u_{i}\,, ~~~~~~ \l^{\a} \rightarrow \rho^{-1} \l^{\a} \,,
}
$$
\bar v^{i} \rightarrow \sigma \bar v^{i}\,, ~~~~~~ \bar\l^{\dot\a} \rightarrow \sigma^{-1} \bar\l^{\dot\a}
$$
where $\rho,\sigma \in {\bf C}$. The two real parameters 
$|\rho|$ and $|\sigma|$ 
are used to impose the normalizations 
$u_{i} \bar u^{i} = 1$ and $v_{i} \bar v^{i} =1$. If one also gauges away 
the overall phases of $u_{i}$ and $\bar v^{i}$, the space 
of harmonic coordinates $u_{i}$ and $\bar v^{i}$ is parametrized by six real parameters. This 
coincides with the number of free parameters of the coset $SU(3) / U(1)\times U(1)$. Indeed, 
we can construct $3 \times 3$ matrices 
$(u^1_i, u^2_i, u^3_i) =(u_i^{(1,0)}, u_i^{(0,-1)}, u_i^{(-1,1)}) $ 
as follows 
\eqn\ide{
u^{1}_{i} \equiv u_i^{(1,0)} = u_{i} \,, ~~~~~
u^{2}_{i} \equiv u_i^{(-1,1)} = \e_{ijk}  \bar v^{j} \bar u^{k}\,, ~~~
u^{3}_{i} \equiv u_i^{(0,-1)} =  v_{i} 
\,. 
}
where $\bar u^{i} = (u_{i})^{*}$ and $v_{i} = (\bar v^{i})^{*}$. Fixing the phases 
of $u^{1}_{i}$ and $u^{3}_{i}$, the $u^{I}_{i}$ form 
$SU(3)$ matrices which are coset representatives of 
${SU(3) \over U(1) \times U(1)}$. 
The $U(1)\times U(1)$ transformations generate 
the phases ${\rm arg}(\rho)$ and ${\rm arg}(\sigma)$. The notation $u_i^{(a,b)}$ indicates the 
$U(1)\times U(1)$ charges of the harmonic variables and they satisfy the  hermiticity property
$\overline{u^{(a,b)}_i} = u^{i(-a,-b)}$. We denote by $u^{i}_I$ the inverse harmonics
\eqn\invers{
u^{i}_I u_{i}^J = \delta_I^{~J}\,, ~~~ u^{I}_i u^j_{ I} = \delta_i^{~j}\,, ~~~~ 
{\rm det} u = \e^{ijk} u^1_i u^2_j u^3_k =1 \,.}
For later use we also list the components of the inverse matrix 
$u^{i}_I$:
\eqn\ideB{
u_{1}^{i} \equiv  u^{i (-1,0)} = \overline{u_i^{(1,0)}} = \bar u^{i} \,, ~~~~~
u_{2}^{i} \equiv u^{i (1,-1)} = \e^{ijk} v_{j} u_{k} \,, ~~~
u_{3}^{i} \equiv u^{i (0,1)} =  \bar v^{i} \,.}

Finally, we consider a further reduction to N=2. We decompose 
the N=3 pure spinors $\l^{\a}_{i}$ and $\bar\l^{\dot \a i}$  into a 
vector of N=2 and a singlet, $\l^{\a}_{i} = (\l^{\a}_{\cal I}, \l^{\a}_{3})$  
and $\bar\l^{\dot \a i} = (\bar \l^{\dot \a {\cal I}},  \bar \l^{\dot \a 3})$ where 
${\cal I} =1,2$.  We set $\l^{\a}_{3}$ and $\bar \l^{\dot \a}_{3}$ to zero. 
The pure spinor equations \reduG\ reduce then to 
\eqn\dueG{
\l^{\a}_{\cal I} \e_{\a\b} \l^{\b}_{\cal J} \e^{\cal I J} = 0\,, ~~~~~~
\bar\l^{\dot\a {\cal J}} \e_{\dot\a\dot\b} \bar\l^{\dot\b {\cal K}} \e_{\cal J K}=0\,,
~~~~~~
\l^{\a}_{\cal I} \bar\l^{\dot\a \cal I} = 0 \,.
} 
The first two equations imply that $\l^{\a}_{\cal I}$ and 
$\bar \l^{\dot \a {\cal I}}$ are factorized into 
$\l^{\a}_{\cal I} = \l^{\a} u_{\cal I}$ and 
$\bar \l^{\dot \a \cal J} = \bar \l^{\dot \a} \bar v^{\cal J}$ 
where $u_{\cal I} \bar v^{\cal I} =0$. 
The vector $\bar v^{\cal I}$ is proportional to $\e^{\cal IJ} u_{\cal J}$. 
Hence without loss of generality one may write 
\eqn\dueH{
\l^{\a}_{\cal I} = \l^{\a} u_{\cal I}\,, ~~~~~
\bar \l^{\dot \a \cal J} = \bar \l^{\dot \a}  \e^{\cal I J}  u_{\cal I}\,.
}
With this parametrization of the N=2 case
there are neither constraints on the $\l$'s nor on the $u$'s. 

The vector $u_{\cal I}$ yields the usual parametrization 
of N=2 harmonic superspace \GalperinUW. 
Namely, one introduces the $SU(2)$ matrix $(u^{+}_{\cal I}, u^{-}_{\cal I})$ 
where $u^{+}_{\cal I} = u_{\cal I}$ and $u^{-}_{\cal I} = (u^{+ {\cal I}})^{*}$ 
with $u^{+}_{\cal J} = \e_{\cal JK} u^{+ {\cal K}}$. The coset 
$SU(2)/U(1)$ is obtained by dividing by the subgroup $U(1)$ which 
generates the phases $u^{\pm}_{\cal I} \rightarrow e^{\pm i \a} u^{\pm}_{\cal I}$.
In fact, eqs. \dueH\ are defined up to a rescaling of 
$\l^\a, \bar\l^{\dot\a}$ and of $u_{\cal I}$ given by 
$u_{\cal I} \rightarrow \rho u_{\cal I}$, for $\rho \neq 0$. 
This yields the compact space ${\bf CP}^{1}$.  


\newsec{N=3 Harmonic Superspace for SYM Theory from Superstrings}

The field equation for $D=4, N=3$ SYM-theory in ordinary (not harmonic) 
superspace are given by 
\lref\soh{
M.F. Sohnius, Nucl. \ Phys. \ B {\bf 136}, 461 (1978);
E.~Witten,  
Phys.\ Lett.\ B {\bf 77}, 394 (1978);   
E.~Witten,  
Nucl.\ Phys.\ B {\bf 266}, 245 (1986).}
\soh
\eqn\reduH{
\{ \nabla^{i}_{\a }, \nabla^{j}_{\b} \} = \e_{\a\b} \bar W^{ij}\,, ~~~~~
\{\bar \nabla_{\dot\a i}, \bar \nabla^j_{\dot\b} \} 
= \e_{\dot\a \dot\b}  W_{ij}\,,
}
$$
\{\nabla^{i}_{\a }, \bar \nabla_{\dot\b j} \} 
= \delta_j^i \nabla_{\a\dot\b} \,.
$$
The coordinates for this N=3 
superspace, $(x^m, \t^{\a}_i, \bar\t^{\dot\a i})$, are obtained by imposing the constraint $\t^{\a}_4 = \bar\t^{\dot\a 4}=0$. Since $\t$'s transform into 
$\l$'s under BRST transformations we also impose for consistency 
$\l^{\a}_{4} =\bar \l^{\dot \a 4} =0$. 

Using 
the decomposition of the N=3 spinors 
$\l^{\a}_{i}$ and $\bar\l^{\dot\a i}$ given in (2.10), 
and contracting the harmonic variables with the operators 
$d_{z \hat\a}$ in \duemezzo\ yields two new spinorial operators 
$$
Q_{G} = \l^{\a} d^{1}_{\a} + \bar\l^{\dot\a} \bar d_{3 \dot \a}\,.
$$
\eqn\harmA{
d^1_\a = u_{i} d_{\a}^{i} = u_{i}^{1} d_\a^i  = u_i^{(1,0)} d_\a^i \,, ~~~~ 
\bar d_{3 \dot\a} = \bar v^{i} \bar d_{\dot \a i} = 
u^{i}_3 \bar d_{\dot\a i} = u^{i (0,1)} \bar d_{\dot \a i}\,. 
}
The operator $d^{1}_{\a}$ 
corresponds to $\xi_{i} D^{i}_{\a}$ and $\bar d_{3 \dot \a}$ to $\eta^{i} \bar D_{\dot \a i}$ in \GalperinUW. 

 Due to the constraints on the $u$'s 
 the operators $d^{1}_{\a}$ and $\bar d_{3 \dot\a}$ satisfy the 
 commutation relations
\eqn\harmAA{
\{d^{1}_{\a}, d^{1}_{\b} \} = 0\,, ~~~~
\{d^{1}_{\a}, \bar d_{3\dot\b} \} = 0\,, ~~~~
\{\bar d_{3\dot\a}, \bar d_{3\dot\b} \} = 0\,.
}
To derive these relations one may use the dimensionally reduced relations 
$\{d_{\a}^{i}, d_{\b}^{j}\} = \e_{\a\b} \Pi^{ij}, \{\bar d_{\dot \a i}, \bar d_{\dot \b j} \} = \e_{\dot \a \dot \b} \bar \Pi_{ij}$ and $\{ d_{\a}^{i}, \bar d_{\dot \a j } \} = \delta^{i}_{j} \Pi_{\a \dot \b}$. 
Hence $Q_{G}$ (where G stands for Grassmann) 
is nilpotent for any $\l^{\a}$ and $\bar\l^{\dot \a}$.

The BRST operator $Q_{G}$ implements naturally the G-analyticity 
on the space of superfields $\Phi(x, \t, \bar\t, \l, \bar\l, u)$. 
A superfield with ghost number zero is given by $\Phi(x, \t, \bar \t, u)$ and G-analyticity 
means $Q_{G} \Phi =0$ which implies $D^{1}_{\a} \Phi = 
\bar D_{3\dot \a} \Phi =0$ (since 
$\{ d^{1}_{\a}, \Phi(x, \t, \bar\t, \l, \bar\l, u)  \} = D^{1}_{\a} \Phi(x, \t, \bar\t, \l, \bar\l, u)$ and similarly for $\bar d_{3 \dot \a}$).
Such a superfield is called a G-analytic  superfield in \GalperinUW. 
A generic superfield $\Phi(x, \t, \bar\t, \l, \bar\l, u)$ with ghost number one  can be parametrized in terms of two $u$-dependent spinorial superfields 
$A_\a,  \bar A_{\dot\a}$ as follows
\eqn\superfield{
\Phi^{(1)}(x, \t, \bar\t, \l, \bar\l, u) = \l^{\a} A_{\a} + \bar\l^{\dot\a} \bar A_{\dot\a}\,,
}
and $\{Q_{G}, \Phi^{(1)} \} =0$ implies the following constraints 
on these superfields
\eqn\harmC{
D^{1}_{\a} A_\b +  D^{1}_{\b} A_\a = 0\,, ~~~~~
\bar D_{3 \dot\a} \bar A_{\dot\b} +  \bar D_{3 \dot\b} \bar A_{\dot \a} = 
0\,, ~~~~~
D^{1}_{\a} \bar A_{\dot\b}  + \bar D_{3 \dot\b}  A_\a  = 0 \,.
}
Assuming that $A_{\a}$ and $A_{\dot\a}$ 
factorize in the same way 
as $D^{1}_{\a} = u_{i} D^{i}_{\a}$ and 
$\bar D_{3\dot\a} = \bar v^{i} \bar D_{\dot \a i}$, so 
$A_{\a} = u_{i} A^{i}_{\a}$ and $A_{\dot \a} = \bar v^{i} A_{\dot \a i}$, 
the equations \harmC\ reproduce (3.1). 
We stress that \harmC, unlike (3.1), do not put the theory on-shell; only 
the extra assumption of the factorization of $A_{\a}$ 
and $A_{\dot \a}$ puts the theory on-shell.  

Gauge transformations are generated by a ghost-number zero scalar 
superfield $\Omega^{(0)}$. To lowest order in $\Phi^{(1)}$ they 
read $\delta \Phi^{(1)} = \{Q_{G}, \Omega^{(0)}\}$ which 
yields $\delta A_{\a} = D_{\a} \Omega$ and 
$\delta A_{\dot \a} = \bar D_{\dot \a} \Omega$. Equations \harmC\ are easily solved in D=4; they 
imply that the superfields $A_{\a}$ and $\bar A_{\dot \a}$ are pure gauge. 
Hence the $Q_{G}$-cohomology in the space of superfields with ghost 
number 1 vanishes. 

To determine on which harmonic variables superfields depend, 
we construct a second BRST operator $Q_{H}$ which is constructed 
from the $SU(3)$ generators 
\eqn\X{
d^{a}_{~~b} = 
u^{a}_{i} \p_{u^{b}_{i}} - u^{i}_{b} \p_{u^{i}_{a}} = 
u^{a}_{i} p^{i}_{b} - u^{i}_{b} p_{i}^{a} \,.
}
where $p_{b}^{i}$ can be represented by $\p/ \p u^{b}_{i}$ and
similarly for $p_{i}^{b}$. 
These generators split into three raising operators 
$d^{1}_{2}=d^{(2,-1)}, 
d^{2}_{3}=d^{(-1,2)}, d^{1}_{3}=d^{(1,1)}$, three lowering operators 
$d_{1}^{2}=d^{(-2,1)}, d^{3}_{2}=d^{(1,-2)}, 
d^{3}_{1}=d^{(-1,-1)}$, and two Cartan 
generators $d^{1}_{1}$ and $d^{2}_{2}$. 
The raising operators operators commute with $Q_{G}$
\eqn\harmF{
[d^{(2,-1)}, d^1_\a] = [d^{(-1,2)}, d^1_\a]=  [d^{(1,1)}, d^1_\a] =0\,,
}
$$
[d^{(2,-1)}, d_{3\dot\a}] = [d^{(-1,2)}, d_{3\dot\a}]=  [d^{(1,1)}, d_{3\dot\a}] =0\,. 
$$
and form an algebra, in particular $[d^{(2,-1)} , d^{(-1,2)} ]= d^{(1,1)}$. 
This suggests to construct a new nilpotent BRST operator  
$Q_{H}$
\eqn\QH{
Q_{H} = \xi^{3}_1  \, d^1_{3} + \xi_{1}^{2}  \, d^1_{2} + 
\xi_{2}^{3} \, d^{2}_{3} - \b^{1}_{3} \xi_{1}^{2} \xi_{2}^{3}  \,, 
}
where we introduced new pairs of anticommuting 
(anti)ghosts $(\xi^{3}_1, \b^{1}_{3})$,  $(\xi^{2}_{1}, \b^{1}_{2})$, 
$(\xi_{2}^{3}, \b^{2}_{3})$  with 
canonical anticommutation relations. 
It is convenient to 
use a notation in which the $U(1) \times U(1)$ weights are made explicit 
$\xi^{3}_1  \equiv \xi^{(-1,-1)}, \xi_{1}^{2} \equiv \xi^{(-2,1)}$ and 
$\xi^{3}_{2} = \xi^{(1, -2)}$. 


Since $Q_{H}$ and $Q_{G}$ anticommute their sum $Q_{tot}$ 
is obviously nilpotent. 
A generic superfield  $\Phi^{(1)}$ 
with ghost number one can be decomposed into  into the following pieces
\eqn\harmH{
\Phi^{(1)} = \l^\a A^{(1,0)}_{\a} +  \bar\l^{\dot\a} \bar A^{(0,1)}_{\dot\a} + 
\xi^{3}_{1}  \, A^{(1,1)} + \xi^{2}_{1}  \, A^{(2,-1)} + \xi^{3}_{2} \, A^{(-1,2)}
}
where $ A^{(1,0)}_{\a}, \bar A^{(0,1)}_{\dot\a}, A^{(2,-1)}, A^{(-1,2)}$ and $A^{(1,1)}$ are 
harmonic superfields (superfields which depend on the variables $u$). 
The harmonic weights of the superfields 
follow from requiring 
that $\Phi^{(1)}$ has zero harmonic weight, just like 
the BRST charge $Q_{tot}$.  Note that 
$\Phi^{(1)}$ depends only upon the variables $x, \t, \bar \t, \l, \bar\l$'s and 
$u$'s and not 
upon the conjugated momenta as a consequence of quantum mechanical rules. 
This forbids ghost-number one combinations of the form $\b \xi \xi, \b \xi \l, \dots$.  

The equations of motion for N=3 SYM follow 
from the BRST-cohomology equations 
\eqn\CS{
\{ Q_{tot}, \Phi^{(1)} \} + {1\over 2} \{ \Phi^{(1)} , \Phi^{(1)} \} =0 \,.
}
Decomposing the superfield $\Phi^{(1)}$ into $\Phi^{(1)}_{H} + \Phi^{(1)}_{G}$, 
where $\Phi^{(1)}_{H}$ denotes the terms with $\xi$-ghosts and 
$\Phi_{G}^{(1)}$ the terms with $\l$-ghosts, the Maurer-Cartan 
equations in \CS\ decompose as follows 
\eqn\CSA{
\{ Q_{G}, \Phi^{(1)}_{G} \} + {1\over 2} \{ \Phi^{(1)}_{G} , \Phi^{(1)}_{G} \} =0 \,,
}
\eqn\CSB{
\{ Q_{G}, \Phi^{(1)}_{H} \} + 
\{ Q_{H}, \Phi^{(1)}_{G} \} + \{ \Phi^{(1)}_{G} , \Phi^{(1)}_{H} \} =0 \,,
}
\eqn\CSC{
\{ Q_{H}, \Phi^{(1)}_{H} \} + {1\over 2} \{ \Phi^{(1)}_{H} , \Phi^{(1)}_{H} \} =0 \,.
}
This system of equations is invariant under the infinitesimal gauge transformation
\eqn\GT{
\Phi^{(1)} \rightarrow \Phi^{(1)} + 
\{Q_{tot}, \Omega\} + \{\Phi^{(1)}, \Omega\}\,,
}
where $\Omega$ is a generic harmonic superfield 
with ghost number zero. According to the above 
decomposition of $\Phi^{(1)}$, one obtains 
$\delta \Phi^{(1)}_{G} = \{Q_{G}, \Omega\} + \{\Phi^{(1)}_{G}, \Omega\}$ 
and $\delta \Phi^{(1)}_{H} = \{Q_{H}, \Omega\} + \{\Phi^{(1)}_{H}, \Omega\}$.  

To reduce the system of equations 
in \CSA-\CSC\ to the field equations of harmonic 
superspace, 
we use the fact that $Q_{G}$ has no cohomology. This implies that 
equation \CSA\ is solved by a pure gauge superfield $\Phi^{(1)}_{G} = e^{-i \Delta} \Big(Q_{G} e^{i \Delta}\Big)$ 
where $\Delta$ is a ghost-number zero superfield known in the 
literature as the {\it bridge} (see for example \GalperinUW). 
Also the BRST cohomology of $Q_{H}$ vanishes on the 
unconstrained superspace and therefore one can also solve the system \CSA-\CSC\ starting from the last equation. 

In the harmonic superspace framework, one usually employs 
the {bridge} superfield $\Delta(x, \t, \bar\t, u)$ to bring the spinorial covariant 
derivatives to the `pure gauge' form 
\eqn\harmI{
\nabla^{(1,0)}_\a = e^{- i\, \Delta}  d^{(1,0)}_\a e^{ i\, \Delta} \,,  ~~~~
\bar\nabla^{(0,1)}_{\dot\a} = e^{- i\, \Delta}  \bar d^{(0,1)}_{\dot\a} e^{ i\, \Delta}\,.  
}
Here the bridge is seen as the most general solution of \CSA. By making a 
finite gauge transformation which sets $\Phi^{(1)}_{G}=0$, 
the gauge transformed $\Phi^{(1)}_{H}$ is given by 
\eqn\REDE{
e^{-i \Delta}(\Phi^{(1)}_{H} +  Q_{H}) e^{i \Delta} = 
\xi^{3}_{1}  \, V^{(1,1)} + \xi^{2}_{1}  \, V^{(2,-1)} + \xi^{3}_{2} \, V^{(-1,2)} \,. 
}
 Equation \CSB\  becomes 
\eqn\EMB{
D^{(1,0)}_\a V^{(2,-1)} =  D^{(1,0)}_\a  V^{(-1,2)} = D^{(1,0)}_\a V^{(1,1)} =0\,,
}
$$
 \bar D^{(0,1)}_{\dot \a} V^{(2,-1)} =   \bar D^{(0,1)}_{\dot \a} V^{(-1,2)} = 
 \bar D^{(0,1)}_{\dot \a} V^{(1,1)} =0\,,
$$
expressing the $G$-analyticity of the harmonic connections 
$V^{(1,1)}, V^{(2,-1)}$ and $V^{(-1,2)}$. The last equation \CSC\ 
finally gives the SYM equations of motion of N=3 harmonic superspace 
$$
D^{(2,-1)} V^{(-1,2)} -  D^{(-1,2)} V^{(2,-1)} + 
\Big[ V^{(2,-1)}, V^{(-1,2)}\Big]
= V^{(1,1)} \,,
$$
$$
D^{(2,-1)} V^{(1,1)} -   D^{(1,1)}V^{(2,-1)} 
+\Big[ V^{(2,-1)}, V^{(1,1)}\Big]
= 0\,,
$$
\eqn\EMC{
D^{(-1,2)} V^{(1,1)} -  D^{(1,1)} V^{(-1,2)} 
+\Big[ V^{(-1,2)}, V^{(1,1)}\Big]
=0\,. 
}
where the harmonic derivatives $D^{(1,1)}, D^{(2,-1)}$ and $D^{(-1,2)}$ 
represent the action of $d^{(1,1)}, d^{(2,-1)}$ and $d^{(-1,2)}$ 
on $u$-dependent superfields. These are the field equations 
of $N=3$ SYM harmonic superspace, see eq.~(12.57) in \GalperinUW. 
Equations \EMB-\EMC\ are invariant under the gauge transformations 
\eqn\harmL{
\delta V^{(2,-1)} = D^{(2,-1)} \omega + \Big[V^{(2,-1)}, \omega\Big]\,, ~~~~~ 
\delta V^{(-1,2)} = D^{(-1,2)} \omega +\Big[V^{(-1,2)}, \omega\Big]\,, ~~~~~ 
}
$$
\delta V^{(1,1)}  =  D^{(1,1)} \omega + \Big[V^{(1,1)}, \omega\Big]\,, 
$$
where the superfield $\omega$ satisfies 
\eqn\harmM{
D^{(1,0)}_\a \omega = 0\,, ~~~~~~~
\bar D^{(0,1)}_{\dot\a} \omega 
= 0\,. 
}


\newsec{The Action and Measure for $N=3$ SYM theory}

We start from the observation that the field equations \CS\ are of
Chern-Simons form and can be derived from an action of the form
\eqn\CSA{
S_{CS} = \int d\mu \Big( \Phi^{(1)} Q_{tot} \Phi^{(1)} + {2\over 3}  \Phi^{(1)} \star 
\Phi^{(1)} \star \Phi^{(1)} \Big)
}
where $\star$ denotes conventional matrix multiplication. 
The measure $d\mu$ has to be determined. 


Instead of dimensionally reducing \CSA\ 
we follow a different path. We have to define the integration measure for all 
zero modes in the theory. Since we are dealing with worldline models, 
the only contribution comes from the zero modes of 
$x^\mu, \t^\a_i, \bar\t^{\dot \a i}$,$ \l^\a_i, \bar\l^{\dot \a i},$ $u_i^I$ 
and  $\xi^1_3, \xi^2_1,\xi^1_2$. The set of ghosts 
$ \l^\a_i, \bar\l^{\dot \a i}$ 
pertains to the BRST 
charge $Q_G$ which implements the $G$-analyticity. 
Therefore, they implement kinematical constraints 
on the theory expressed by the equations:
\eqn\Gana{
[Q_G, S_{N=3}] =0\,, \quad\quad
[Q_G, d\mu_H] =0\,, 
}
where $S_{N=3}$ is the off-shell $N=3$ action and 
$d\mu_H$ is the invariant measure in the space of the zero modes 
of $x^\mu, \t^\a_i, \bar\t^{\dot \a i}, u_i^I$ and  $\xi^1_3, \xi^2_1,\xi^1_2$. In addition, 
$S_{N=3}$ has zero ghost number, while $ d\mu_H$ has ghost number 
three. Form \berko\ and 
\lref\BerkovitsGJ{
N.~Berkovits, M.~T.~Hatsuda and W.~Siegel,
Nucl.\ Phys.\ B {\bf 371}, 434 (1992)
[hep-th/9108021].
} \BerkovitsGJ it is known that $d\mu_H \in H^3(Q_H)$. This implies that 
$d\mu_H = d\xi^1_3 d\xi^2_1 d\xi^1_2 d\mu'$ where 
the measure $d\mu' = d\mu'(x^\mu, \t^\a_i, \bar\t^{\dot \a i}, u_i^I)$ has to be 
fixed by the G-analyticity \Gana. 

First we consider the space formed by  $x^\mu, \t^\a_i, \bar\t^{\dot \a i}$. 
The conditions in \Gana\ select the analytic subspace
$(x^m_A, \t^{(0,1)}_\a, \t^{(1,-1)}_\a$, $
 \bar\t^{(1,0)}_{\dot\a},  \bar\t^{(-1,1)}_{\dot\a})$ where 
 $\t^{(a,b)} = u^{i (a,b)} \t_{i}$, and 
 $x_{A}^{\a\dot\a} = x^{\a\dot\a} + 2 i \t^{\a (-1,0)} \bar\t^{\dot\a(1,0)} 
 + 2i \t^{\a (0,1)} \bar\t^{\dot \a (0,-1)}
 $ 
 Therefore the only invariant measure is given by 
\eqn\GanaB{
d\mu' = d^4x_A d^{2}\t^{(0,1)} d^{2}\t^{(1,-1)} d^{2}\bar\t^{(1,0)}  
d^{2}\bar\t^{(-1,1)} d\mu_u
} 
where $d\mu_u$ is the measure for the harmonic variables. 
In order to derive a $Q_G$ invariant measure $d\mu_u$, 
we introduce the new variables (projective harmonic variables 
\lref\sc{
A.~A.~Roslyi and A.~S.~Schwarz,
Commun.\ Math.\ Phys.\  {\bf 105}, 645 (1986).
}
\sc) 
\eqn\newVAE{
z_1 = u_1/u_3\,, \quad\quad 
z_2 = u_2/u_3\,, \quad\quad
z_3 = v_1/v_2\,. 
}

The three raising and three lowering operators
are three Lie derivatives whose 
duals are six one-forms whose product gives the integration 
measure on $SU(3)/U(1)\times U(1)$. This is the 
Haar measure for  $SU(3)/U(1)\times U(1)$ given by \sc 
\eqn\dmuB{
d\mu_{u} ={ \prod_{i=1}^3 dz_i d\bar z_i   \over 
( 1 + |z_1|^2 + |z_2|^2 )^4 ( 1 + |z_3|^2 + |z_2 +z_{1} z_{3}|^2 )^2\,.
}}


\newsec{Acknowledgemnets}

We thank N. Berkovits, M. Porrati, G. Policastro, 
M. Ro\v cek and W. Siegel for useful discussions. This work was 
partly funded by NSF Grants PHY-0098527. 
PAG thanks L. Castellani and A. Lerda for discussions and financial support. 

 
\listrefs
\bye